# Tethered Particle Motion technique in crowded media: Compaction of DNA by globular macromolecules

Marc JOYEUX [*]

*Laboratoire Interdisciplinaire de Physique,*

*CNRS and Université Grenoble Alpes,*

*38400 St Martin d'Hères,*

*France*

Corresponding Author

*(M.J.) E-mail: marc.joyeux@univ-grenoble-alpes.fr.

ORCID : Marc Joyeux: 0000-0002-6282-1846

**ABSTRACT**

Tethered Particle Motion (TPM) is a single molecule technique, which consists in tracking the motion of a nano-particle (NP) immersed in a fluid and tethered to a glass surface by a DNA molecule. The present work addresses the question of the applicability of TPM to fluids which contain crowders at volume fractions ranging from that of the nucleoid of living bacteria (around 30%) up to the jamming threshold (around 66%). In particular, we were interested in determining whether TPM can be used to characterize the compaction of DNA by globular crowders. To this end, extensive Brownian Dynamics simulations were performed with a specifically built coarse-grained model. Analysis of the simulations reveals several effects not observed in dilute media, which impose constraints on the TPM set-up. In particular, the Tethered Fluorophore Motion (TFM) technique, which consists in replacing the NP by a much smaller fluorophore, is probably better suited than standard TPM. Moreover, a sample preparation technique which does not involve hydrophilic patches may be required. Finally, the use of a DNA brush may be needed to achieve DNA concentrations close to in vivo ones.

# INTRODUCTION

The Tethered Particle Motion (TPM) technique consists in tracking the motion of a nano-particle (NP) immersed in a fluid and tethered to a glass surface by a DNA molecule (or eventually another biopolymer). The size of the NP ranges from tens to hundreds of nm and that of the DNA molecule from several hundreds to few thousands of base pairs. Conformational changes of the DNA induced by changes in the fluid medium or by interactions of the DNA with other molecules in the medium modify the amplitude of the Brownian motion of the tethered particle. Measurement of these variations with a simple confocal optical microscope allows one to quantify the conformational changes of the DNA molecule. This single-molecule technique has been introduced in the early 1990s [1,2] and has recently been enriched along two directions. The first improvement consists in monitoring simultaneously the displacements of several hundreds of tethered NPs positioned in a controlled manner by soft nano-lithography. This technique, called high-throughtput TPM (htTPM), reduces dramatically the acquisition time and gives access to highly refined statistics [3]. The second variant of TPM, called Tethered Fluorophore Motion (TFM), consists in replacing the NP by a much smaller fluorophore bound to the free end of the DNA chain [4,5]. This allows to combine TPM with fluorescence techniques like Förster resonance energy transfer, however at the cost of a significant reduction in observation time due to fluorophore photobleaching.

In the last decades, the TPM technique has been used to investigate several biological mechanisms at play in cells, including the processivity of RNA polymerases [1,2,6] and helicases [7]; the branch migration of individual Holliday junctions [8]; the formation of DNA loops caused by the binding of Lac repressors [9-12] and restriction enzymes [13]; the assembly of the transposome [14]; the action of recombinases and translocases [4,5,15-17]; the bending of DNA by DNA-binding proteins [18,19] and its wrapping around histones [20] and other protein wheels [21]; condensation of DNA by DNA-bridging proteins [22,23]; the influence of DNA supercoiling on the efficiency of an epigenetic switch [24]; the effect of crowding agents on the architectural properties of HU nucleoid proteins [25]. In all these studies, the macromolecular species are added to the medium at relatively low concentration and interact specifically with the DNA but only marginally with the NP. As a consequence, it is sufficient to clean experimental data from spurious points, subtract experimental drift and correct for the blurring effect to properly deduce the apparent length of the DNA molecule from the amplitude of the Brownian motion of the NP [26].

The present work deals instead with the applicability of the TPM technique to media which contain crowders at volume fractions ranging from that in the nucleoid of living bacteria (around 30% [27,28]) up to the jamming threshold (around 66% [29]), that is, the concentration at which macromolecules are so tightly packed that they can no longer move. Indeed, bacterial cells contain about 300 mg/ml of globular macromolecules, which increases their viscosity up to six times that of water [30]. It has been suggested that this large concentration of globular macromolecules may be responsible for the formation of the bacterial nucleoid, that is, for the fact that bacterial DNA does not spread over the whole cell but occupies instead only a fraction thereof [31-35]. The proposed explanation is that a segregative phase separation takes place in the cytoplasm, which leads to a phase rich in DNA but depleted in macromolecules (the nucleoid) and a second phase enriched in macromolecules but depleted in DNA (the cytosol) [35-46]. Moreover, it is known that at even larger crowder concentrations, close to the jamming threshold, long DNA molecules which evolve freely in a solvent collapse abruptly to very dense globules [39], presumably because of a similar segregative phase separation [36,37].

The purpose of the present paper is to report on simulation work, which aims at estimating whether the TPM technique can be used to investigate heavily crowded media and, in particular, the mechanism of DNA compaction by globular crowders. Extensive Brownian Dynamics (BD) simulations were performed with a coarse-grained model adapted from those developed previously to investigate the compaction of DNA molecules by macromolecular crowders [42-46]. Analysis of the simulations reveal several potential problems, which set constraints on TPM set-ups for heavily crowded media.

**MATERIALS AND METHODS**

As is illustrated in Fig. 1, the coarse-grained model used in Brownian Dynamics simulations is composed of a DNA chain and $C$ spherical crowders. The DNA chain consists of $n - 1 = 999$ beads of radius $R_k = R_{\text{DNA}} = 1.0$ nm located at position $\mathbf{r}_k$ ($1 \leq k \leq n - 1$), which are connected by springs and separated at equilibrium by a distance $l_k^0 = l_0 = 2.5$ nm ($1 \leq k \leq n - 2$). Each bead represents 7.5 base pairs (bp), so that the chain represents a DNA molecule containing slightly less than 7500 bp, which is about the maximum DNA length that is used in today's TPM experiments. One end of the DNA chain (bead $k = 1$) is grafted at the origin O of the bottom horizontal plane and the other end (bead $k = n - 1$) is attached to the NP, which is modeled as a sphere of radius $R_n = R_{\text{NP}} = 1$, 20, or 150 nm located at position $\mathbf{r}_n$. Vector $\mathbf{r}_n$ is decomposed into $\mathbf{r}_n = \mathbf{r}_{||} + \mathbf{r}_z$, where $\mathbf{r}_{\parallel}$ is the component of $\mathbf{r}_n$ parallel to the bottom horizontal plane and $\mathbf{r}_z$ the component perpendicular to this plane. The NP is connected to DNA bead $n-1$ by a spring and separated from it by a distance $l_{n-1}^0 = R_{\text{DNA}} + R_{\text{NP}} + 0.5$ nm at equilibrium. Since $l_0 = 2R_{\text{DNA}} + 0.5$ nm, this choice ensures that attaching a NP with radius $R_{\text{NP}} = 1$ nm at the free end of the DNA chain actually amounts to investigating a homogenous DNA chain with $n = 1000$ beads. The $C$ spherical crowders are located at positions $\mathbf{s}_j$ ($1 \leq j \leq C$) and have a radius $R_C = 40$ nm. Except for simulations without crowders ($C = 0$), the DNA chain and the crowders are enclosed in the volume delimited by the bottom horizontal plane and a hemisphere of center O and radius $R_H = 1000$ nm. Simulations were performed with $C = 0$, 1000, 2000, 3000, 3500 or 4000 crowders, corresponding to volume fractions $\rho$ equal to 0.00, 0.13, 0.26, 0.38, 0.45 or 0.51, respectively.

The total potential energy of the system, $E_{\text{pot}}$, is the sum

$$E_{\text{pot}} = E^{\text{DNA}} + \sum_{j=1}^{C} E^{\text{DNA}/j} + \sum_{j=1}^{C-1}\sum_{i=j+1}^{C} E^{j/i} + E^{\text{wall}}, \qquad (1)$$

where $E^{\text{DNA}}$ is the internal energy of the DNA chain (including the terminal NP), $E^{\text{DNA}/j}$ the repulsive electrostatic interaction between the DNA chain (including the terminal NP) and spherical crowder $j$, $E^{j/i}$ the repulsive electrostatic interaction between spherical crowders $j$ and $i$, and $E^{\text{wall}}$ the repulsive potential that maintains all particles inside the confinement volume. The internal energy of the DNA chain is itself the sum of three terms

$$E^{\text{DNA}} = \frac{h}{2}\sum_{k=1}^{n-1}(l_k - l_k^0)^2 + \frac{g}{2}\sum_{k=2}^{n-2}\theta_k^2 + q^2 \sum_{k=1}^{n-4}\sum_{m=k+4}^{n} H(\|\mathbf{r}_k - \mathbf{r}_m\| - R_k - R_m), \qquad (2)$$

where $l_k = \|\mathbf{r}_{k+1} - \mathbf{r}_k\|$ denotes the distance between two successive beads and $\theta_k$ the angle between vectors $\mathbf{r}_k - \mathbf{r}_{k-1}$ and $\mathbf{r}_{k+1} - \mathbf{r}_k$. The first term in the right-hand side of Eq. (2) describes the stretching energy of the DNA chain, the second one its bending energy, and the last one the electrostatic repulsion between different DNA segments. The stretching force constant was set to $h = 100\, k_B T/l_0^2$, where $T = 295$ K, because it ensures that the fluctuations of the distance between neighboring beads remain small enough for the integration time step used in the simulations [47]. The bending force constant was deduced from the known persistence length of DNA, $\xi = 50$ nm, according to $g = \xi\, k_B T/l_0 = 20\, k_B T$. Note that the NP rotates freely around DNA bead $n-1$. Finally, function $H(r)$ is defined according to

$$H(r) = \frac{1}{4\pi\varepsilon r}\exp\left(-\frac{r}{r_D}\right), \qquad (3)$$

where $\varepsilon = 80\varepsilon_0$ denotes the dielectric constant of the buffer and $r_D = 1.07$ nm the Debye length inside the buffer. This value corresponds to a total concentration of monovalent salts of 100 mM. $q$ is the value of the electrostatic charge which is placed at the centre of each DNA bead. It was set to $q = -3.525\bar{e}$, where $\bar{e}$ denotes the absolute charge of an electron, in agreement with Manning's counterion condensation theory [48,49].

For the sake of simplicity, the same electrostatic charge $q$ was placed at the center of all spherical crowders and it was assumed that the repulsion between DNA beads and spherical crowders, as well as the repulsion between two spherical crowders, are governed by the same type of hard-core potential as in Eq. (2), that is

$$E^{DNA/j} = q^2 \sum_{k=1}^{n} H(\|\mathbf{r}_k - \mathbf{s}_j\| - R_k - R_C) \tag{4}$$

and

$$E^{j/i} = q^2 H(\|\mathbf{s}_j - \mathbf{s}_i\| - 2R_C) . \tag{5}$$

Finally, the confining potential is written in the form

$$E^{wall} = 1000\, k_B T \left( \sum_{k=1}^{n} F(\|\mathbf{r}_k\|, R_k) + \sum_{j=1}^{C} F(\|\mathbf{s}_j\|, R_C) \right) + \frac{h}{2} \left( \sum_{k=1}^{n} G(\mathbf{r}_k \cdot \mathbf{z}, R_k) + \sum_{j=1}^{C} G(\mathbf{s}_j \cdot \mathbf{z}, R_C) \right) \tag{6}$$

where $\mathbf{z}$ is the vertical unitary vector and functions $F(r, \sigma)$ and $G(z, \sigma)$ are defined according to if $r \leq R_H - \sigma : F(r, \sigma) = 0$,

if $r > R_H - \sigma : F(r, \sigma) = \frac{r^6}{(R_H - \sigma)^6} - 1 \tag{7}$

and

if $z \leq \sigma : G(z, \sigma) = (z - \sigma)^2$,

if $z > \sigma : G(z, \sigma) = 0 . \tag{8}$

The dynamics of the model was investigated by integrating numerically overdamped Langevin equations. Practically, the new position vector for each particle (DNA bead or spherical crowder), $\mathbf{x}_j^{(i+1)}$, was computed from the current position vector, $\mathbf{x}_j^{(i)}$, according to

$$\mathbf{x}_j^{(i+1)} = \mathbf{x}_j^{(i)} + \frac{D_j}{k_B T} \Delta t\, \mathbf{F}_j^{(i)} + \sqrt{2 D_j\, \Delta t}\, \boldsymbol{\zeta}_j^{(i)} , \tag{9}$$

where the translational diffusion coefficient $D_j$ is equal to $(k_B T)/(6\pi\eta R_j)$ for DNA beads and to $(k_B T)/(6\pi\eta R_C)$ for spherical crowders. $\eta = 0.00089$ Pa s is the viscosity of the buffer at $T = 298$ K. $\mathbf{F}_j^{(i)}$ is the vector of inter-particle forces arising from the potential energy $E_{pot}$ and $\boldsymbol{\zeta}_j^{(i)}$ a vector of random numbers extracted at each step $i$ from a Gaussian distribution of mean 0 and variance 1. $\Delta t$ is the integration time step, which was set to 10.0 ps.

Simulations were performed as follows. The DNA chain (including the terminal NP) was first equilibrated for 2 seconds, either with (for $C \neq 0$) or without (for $C = 0$) the confinement hemisphere. For $C \neq 0$, the crowders were then introduced at random, homogenously distributed and non-overlapping positions and the DNA/crowders system was equilibrated again for 500 ms (for $C \leq 3500$) or 700 ms (for $C = 4000$). The production step then began. Six simulations with different initial conditions and different sets of random numbers were run for each pair of values of $C = 0$, 1000, 2000, 3000, 3500 or 4000 and $R_{NP} = 1$, 20 or 150 nm. Depending on the value of $C$, the total integration time for each pair $(C, R_{NP})$ ranged from 2 to 6 seconds, that is, from 10 to 60 times the NP lateral displacement relaxation times obtained from the two-time correlation function according to Eq. (3) of [26]. Quantities of interest were sampled every $10^7$ steps (0.1 ms) and averaged. Fig. 1 shows a typical equilibrated conformation obtained with $C = 3000$ crowders and a NP of radius $R_{NP} = 150$ nm.

Several comments are in order here.

First, investigation of DNA compaction requires that DNA molecules much longer than its persistence length $\xi = 50$ nm be used. This is the reason, why DNA molecules with 7500 bp ($50\xi$) were used in the present simulations, instead of the shorter molecules in the range 300-3000 bp that are most commonly used in TPM experiments.

Moreover, the radius of the spherical crowders used in the simulations is about 10 times larger than the radii of PEG polymers (radius of gyration of PEG 8000 is about 2.5 nm at very small polymer concentrations and salt concentrations ranging from 10 mM to 200 mM [50]) and BSA proteins (hydrodynamic radius ≈ 4.5 nm at pH ≈7 [51]) used in many experiments with globular crowders. Such a large value of $R_C$ is mandatory to keep the number of crowders small enough to allow for integration times of the order of several seconds for each value of the radius $R_{NP}$ and each concentration of crowders. This choice is still meaningfull, because the compaction of the DNA coil is governed by the effective volume fraction of crowders [42], not their radius, at least as long as this radius remains small compared to the radius of gyration of the DNA coil.

Finally, it is worth noting that the radius $R_\text{H} = 1000$ nm of the confinement hemisphere is significantly smaller than the length of the fully extended DNA chain (2500 nm). However, $R_\text{H}$ is approximately twice as large as the average end-to-end distance of coiled DNA, $(2nl_0/\xi)^{1/2} \approx 500$ nm. Moreover, in simulations performed without crowders ($C = 0$) and without confinement hemisphere, NPs with radius $R_\text{NP} = 1$ nm spend only 0.12% of the time at a distance from the origin O larger than 999 nm and NPs with radius $R_\text{NP} = 20$ nm only 1.5% of the time at a distance larger than 980 nm. This explains why NPs with $R_\text{NP} = 1$ or 20 nm are almost never in contact with the confinement hemisphere in simulations with $C \neq 0$. In contrast, NPs with radius $R_\text{NP} = 150$ nm spend about 18% of the time at a distance from the origin O larger than 850 nm in simulations with $C = 0$. The confinement hemisphere may therefore affect to a small extent the results obtained in simulations with $R_\text{NP} = 150$ nm and $C \neq 0$.

**RESULTS**

**Compaction of the DNA coil with increasing concentration of crowders**

Compaction of a long DNA molecule immersed in a media containing globular crowders is governed by the effective volume fraction of crowders, $\rho_\text{e}$ [42]. For the geometry of the TPM coarse-grained model and the repulsive potential in Eq. (5), $\rho_\text{e}$ is related to the number of crowders, $C$, and the volume fraction of crowders, $\rho$, according to

$$\rho_\text{e} = 2C \left(\frac{R_\text{C}+\Delta R_\text{C}}{R_\text{H}}\right)^3 = \rho \left(\frac{R_\text{C}+\Delta R_\text{C}}{R_\text{C}}\right)^3 , \qquad (10)$$

where $2(R_\text{C} + \Delta R_\text{C})$ is the distance at which the repulsion energy between two spherical crowders is equal to $k_\text{B}T$ [42]. For the repulsion potential in Eq. (5), $\Delta R_\text{C}$ is equal to 0.865 nm. $C = 0$, 1000, 2000, 3000, 3500 and 4000 correspond therefore to $\rho_\text{e} = 0.00$, 0.14, 0.27, 0.41, 0.48, and 0.55, respectively (remember that hard spheres become jammed at $\rho_\text{e} \approx 0.66$ [29]).

Compaction of the DNA chain was quantified by the evolution of the mean radius of gyration of the DNA coil, $\langle R_\text{g}^2 \rangle^{1/2}$, and the mean end-to-end distance of the DNA molecule, $\langle \mathbf{R}_\text{ee}^2 \rangle^{1/2}$, which were computed according to

$$\langle R_g^2\rangle^{1/2} = \langle\frac{1}{2(n-1)^2}\sum_{k=1}^{n-1}\sum_{m=1}^{n-1}\|\mathbf{r}_k - \mathbf{r}_m\|^2\rangle^{1/2} \tag{11}$$

and

$$\langle \mathbf{R}_{ee}^2\rangle^{1/2} = \langle\|\mathbf{r}_{n-1} - \mathbf{r}_1\|^2\rangle^{1/2}, \tag{12}$$

respectively. The values of $\langle R_g^2\rangle^{1/2}$ and $\langle \mathbf{R}_{ee}^2\rangle^{1/2}$ obtained from BD simulations performed with different values of $\rho_e$ and $R_{NP}$ are displayed as scattered symbols in Fig. 2. For each value of $R_{NP}$, both $\langle R_g^2\rangle^{1/2}$ and $\langle \mathbf{R}_{ee}^2\rangle^{1/2}$ appear to decrease linearly with $\rho_e$ between $\rho_e = 0$ and $\rho_e = 0.41$. Least squares fits were performed to determine the function

$$f(\rho_e) = A + B\rho_e \tag{13}$$

which best reproduces the four points with $0 \leq \rho_e \leq 0.41$ for each value of $R_{NP}$. Fitted coefficients $A$ and $B$ are shown in Table 1 and corresponding functions $f(\rho_e)$ are plotted in Fig. 2 as dot-dashed lines. Finally, the relative decrease of each quantity between $\rho_e = 0$ and $\rho_e = 0.41$ was estimated according to

$$\Delta = \frac{f(0.41) - f(0)}{f(0)}. \tag{14}$$

Computed values of $\Delta$ are also shown in Table 1.

Fig. 2 furthermore indicates that the points with $\rho_e > 0.41$ fall instead off the straight lines determined from smaller values of $\rho_e$. For $R_{NP} = 1$ nm and $\rho_e = 0.55$, both $\langle R_g^2\rangle^{1/2}$ and $\langle \mathbf{R}_{ee}^2\rangle^{1/2}$ decrease by about 45% compared to $\rho_e = 0.41$, which reflects a sharp enhancement of the compaction of the DNA coil above $\rho_e = 0.41$. Simulations with $R_{NP} = 20$ and 150 nm were performed at $\rho_e = 0.48$ instead of $\rho_e = 0.55$, because at $\rho_e = 0.55$ the dynamics of the NPs is too slow to get converged results in a reasonable amount of time. Corresponding points in Fig. 2 also denote a somewhat enhanced compaction of the DNA coil compared to smaller values of $\rho_e$, but the effect is much less pronounced than for $R_{NP} = 1$ nm and $\rho_e = 0.55$.

Both regimes, the slow linear compaction of the DNA molecule below $\rho_e = 0.41$ as well as the more abrupt compaction above $\rho_e = 0.41$, are analyzed in detail in Section Discussion below.

**Reduction of the mean displacement of the NP with increasing concentration of crowders**

The signal recorded in TPM experiments is the quadratic mean of the lateral displacements of the NP, which in the simulations corresponds to $\langle \mathbf{r}_{\parallel}^2 \rangle^{1/2}$. The values of $\langle \mathbf{r}_{\parallel}^2 \rangle^{1/2}$ obtained from BD simulations performed with different values of $\rho_e$ and $R_{NP}$ are displayed as scattered symbols in the upper plot of Fig. 3. In addition, the bottom plot of Fig. 3 displays the values of $\langle \mathbf{r}_{z,b}^2 \rangle^{1/2} = \langle (\mathbf{r}_z - R_{NP}\mathbf{z})^2 \rangle^{1/2}$, that is, the evolution with $\rho_e$ of the quadratic mean of the vertical excursion of the bottom of the NP. This quantity is of interest, because theory predicts that a big solid sphere immersed in a sea of smaller ones should localize close to the bounding wall [45,52]. It is consequently necessary to check whether NPs used in TPM experiments are subject to this effect.

Comparison of Figs. 2 and 3 indicates that the compaction of the DNA coil and the motion of the NP are highly correlated. In particular, the evolution of $\langle \mathbf{r}_{\parallel}^2 \rangle^{1/2}$ and $\langle \mathbf{r}_{z,b}^2 \rangle^{1/2}$ in the range $0 \leq \rho_e \leq 0.41$ reflects the linear decrease of $\langle R_g^2 \rangle^{1/2}$ and $\langle \mathbf{R}_{ee}^2 \rangle^{1/2}$ observed in Fig. 2. As for $\langle R_g^2 \rangle^{1/2}$ and $\langle \mathbf{R}_{ee}^2 \rangle^{1/2}$, least squares fits were performed to determine the linear functions which best reproduce the values of $\langle \mathbf{r}_{\parallel}^2 \rangle^{1/2}$ and $\langle \mathbf{r}_{z,b}^2 \rangle^{1/2}$ in the range $0 \leq \rho_e \leq 0.41$ for each value of $R_{NP}$. Fitted coefficients $A$ and $B$ and relative decreases $\Delta$ between $\rho_e = 0$ and $\rho_e = 0.41$ are shown in Table 1. Adjusted linear functions $f(\rho_e)$ are furthermore plotted as dot-dashed lines in Fig. 3.

Moreover, the abrupt reduction of $\langle \mathbf{r}_{\parallel}^2 \rangle^{1/2}$ by 43% for $R_{NP} = 1$ nm and $\rho_e = 0.55$ with respect to $\rho_e = 0$ compares well with the reduction of $\langle R_g^2 \rangle^{1/2}$ and $\langle \mathbf{R}_{ee}^2 \rangle^{1/2}$ by 45% in the same range. In contrast, simulations reveal a marked difference between the dynamics of NPs with radius $R_{NP} = 150$ nm compared to $R_{NP} = 1$ or 20 nm above $\rho_e = 0.41$. Indeed, for $\rho_e = 0.48$, NPs with radius $R_{NP} = 150$ nm remain in the vicinity of the bottom plane (more precisely, in the vicinity of the first layer of crowders that pave the bottom plane) for most of the ulterior time steps, as soon as they are first brought in its neighborhood by random thermal motion, in agreement with theoretical predictions [52]. This is clearly seen in the bottom plot of Fig. 3 where, for $R_{NP} = 150$ nm, $\langle \mathbf{r}_{z,b}^2 \rangle^{1/2}$ drops abruptly down to 65 nm at $\rho_e = 0.48$, which means that the bottom of the NP is located on average only 25 nm above the radius of spherical crowders. NPs with radius $R_{NP} = 150$ nm are indeed significantly larger than crowders with radius $R_C = 40$ nm, and the size difference is sufficient to let the NP localize in the neighborhood of the bottom plane [52]. In contrast, Fig. 3 indicates that NPs with $R_{NP} = 20$ nm do not remain close to the bottom plane at $\rho_e = 0.48$, which is due to the fact that, in this case, crowders are bigger than the NP.

The two regimes, $\rho_e \leq 0.41$ and $\rho_e > 0.41$, are discussed in detail in Section Discussion.

**DISCUSSION**

As mentioned in Introduction, the purpose of the present work was to estimate whether the TPM technique can be used to investigate heavily crowded media and, in particular, the mechanism of DNA compaction by globular crowders. In the present section, we discuss how the extensive BD simulations described above help answer this question. In particular, we highlight several potential problems and propose solutions to overcome them.

**Perturbations imposed to the DNA coil by the TPM setup**

Adjusted parameters $A$ (Table I) represent a good estimate of a given quantity in the absence of crowders and can be used to check the strength of the perturbations imposed to the DNA coil by the TPM set-up itself. For example, the Worm-Like-Chain (WLC) model predicts that the mean radius of gyration of a free polymer with contour length $(n-2)l_0$ and persistence length $\xi$ is $\langle R_g^2 \rangle^{1/2} = ((n-2)l_0\xi/3)^{1/2}$ [53], that is about 203.9 nm for $n = 1000$, $l_0 = 2.5$ nm and $\xi = 50$ nm. According to Table 1, the corresponding values obtained from BD simulations are 204.1 nm for $R_{NP} = 1$ nm, 207.9 nm for $R_{NP} = 20$ nm, and 218.1 nm for $R_{NP} = 150$ nm. The TPM set-up has consequently a negligible influence on the mean radius of gyration of the DNA coil for $R_{NP} = 1$ and 20 nm, and is responsible for an increase of $\langle R_g^2 \rangle^{1/2}$ smaller than 7% for $R_{NP} = 150$ nm. The WLC model also predicts that the mean end-to-end distance of the free DNA chain is $\langle \mathbf{R}_{ee}^2 \rangle^{1/2} = (2(n-2)l_0\xi)^{1/2} \approx 499.5$ nm [53], whereas BD simulations lead to 544.3 nm for $R_{NP} = 1$ nm, 562.5 nm for $R_{NP} = 20$ nm, and 635.7 nm for $R_{NP} = 150$ nm, that is, increases of about 9%, 13% and 27%, respectively, for the TPM set-up compared to the free DNA chain. The significant increase (+9%) observed for homogeneous DNA chains ($R_{NP} = 1$ nm) is probably due to the fact that the TPM set-up imposes that the end of the DNA chain grafted to the bottom surface remains outside (or at the outer surface) of the DNA coil. Moreover, the large discrepancy (+27%) observed for NPs with $R_{NP} = 150$ nm is probably due to the fact that for this geometry the other end of the DNA chain, which is grafted to the NP, is also constrained to remain outside (or at the outer surface) of the DNA coil, because NPs are actually too big to penetrate inside the DNA coil.

**Correlation between the size of the DNA coil and the lateral displacement of the NP**

As already mentioned in Section Results, the properties of the DNA coil and the displacements of the NP are highly correlated in the first linear regime extending up to $\rho_e \approx 0.41$.

This correlation can be rationalized theoretically. Indeed, in [54], Segall *et al* provided a theoretical analysis of the motion of the NP in TPM experiments. Based, in particular, on the fact that there exist only two relevant length scales in the experiment, namely the radius $R_{\text{NP}}$ of the NP and the radius of gyration $(L\xi/3)^{1/2}$ of the isolated DNA coil, they derived expressions for $\langle \mathbf{r}_{||}^2 \rangle$ and $\langle \mathbf{r}_z^2 \rangle$ in terms of the radius of gyration of the DNA coil and of the excursion number $N_R$, which is defined as the ratio of the two relevant length scales (Eqs. (10b) and (10c) of [54]). Their work dealt with a standard TPM experiment without crowders, but the results of simulations performed with spherical crowders may eventually be compared to their theoretical predictions by simply replacing the radius of gyration of the isolated DNA coil, $(L\xi/3)^{1/2}$, by its actual radius of gyration $\langle R_g^2 \rangle^{1/2}$. This amounts to defining the excursion number according to

$$N_R = \frac{R_{\text{NP}}}{\langle R_g^2 \rangle^{1/2}} \tag{15}$$

and recasting Eqs. (10b) and (10c) of [54] in the form

$$\frac{\langle \mathbf{r}_{||}^2 \rangle}{\langle R_g^2 \rangle} = 2 + \frac{4 N_R}{\sqrt{\pi}\, \text{erf}(N_R)}, \tag{16}$$

$$\frac{\langle \mathbf{r}_z^2 \rangle}{\langle R_g^2 \rangle} = 2 + \frac{4 N_R}{\sqrt{\pi}\, \text{erf}(N_R)} + N_R^2. \tag{17}$$

The plots of $\langle \mathbf{r}_{||}^2 \rangle / \langle R_g^2 \rangle$ and $\langle \mathbf{r}_z^2 \rangle / \langle R_g^2 \rangle$ as a function of $N_R$ obtained from BD simulations, as well as the theoretical estimates in the right-hand sides of Eqs. (16) and (17), are shown in Fig. 4. Comparison of Fig. 4 with Fig. 2 of [54] indicates that the agreement between simulations and theoretical results is very similar for simulations with and without crowders. More precisely, the agreement is very good for $\langle \mathbf{r}_{||}^2 \rangle / \langle R_g^2 \rangle$, whereas the theoretical estimate of $\langle \mathbf{r}_z^2 \rangle / \langle R_g^2 \rangle$ is somewhat too large for values of $N_R$ smaller than 1. The only exception is the point obtained from simulations with $R_{\text{NP}} = 150$ nm and $\rho_e = 0.48$, whose vertical component clearly falls off the theoretical curve, because of the localization of the NP close to the bottom surface discussed above. It is noted that the excursion number $N_R$ is close to 1 for a 7500 bp DNA and a NP with radius $R_{\text{NP}} = 150$ nm, which indicates that the system is at the boundary between molecule-dominated motion ($N_R < 1$) and NP-dominated motion ($N_R > 1$, confined rotations).

**Analysis of the linear regime at intermediate crowder concentrations**

The major information derived from Table I is however that in the range $0 \leq \rho_e \leq 0.41$ all quantities decrease much more rapidly as a function of $\rho_e$ for $R_{NP} = 20$ and 150 nm than for $R_{NP} = 1$ nm. More precisely, $\langle R_g^2 \rangle^{1/2}$, $\langle \mathbf{R}_{ee}^2 \rangle^{1/2}$, $\langle \mathbf{r}_\parallel^2 \rangle^{1/2}$ and $\langle \mathbf{r}_{z,b}^2 \rangle^{1/2}$ decrease by only a few percents for $R_{NP} = 1$ nm, whereas the drop is of the order of 20% for $R_{NP} = 20$ and 150 nm. This is all the more unexpected, as the computed persistence length of the DNA chain decreases from 52.0 nm at $\rho_e = 0$ down to 46.6 nm at $\rho_e = 0.41$ for $R_{NP} = 1$ nm, from 52.3 nm to 46.2 nm for $R_{NP} = 20$ nm, and from 53.2 nm to 46.5 nm for $R_{NP} = 150$ nm. Stated in other words, the persistence length of the DNA chain decreases by about 12% between $\rho_e = 0$ and $\rho_e = 0.41$, whatever the radius of the NP. According to the WLC model, this corresponds to a decrease of $\langle R_g^2 \rangle^{1/2}$ and $\langle \mathbf{R}_{ee}^2 \rangle^{1/2}$ by about 6%, which is consistent with the results obtained for $R_{NP} = 1$ nm but significantly smaller than the drops measured for $R_{NP} = 20$ and 150 nm.

The steeper decrease of $\langle R_g^2 \rangle^{1/2}$, $\langle \mathbf{R}_{ee}^2 \rangle^{1/2}$, $\langle \mathbf{r}_\parallel^2 \rangle^{1/2}$ and $\langle \mathbf{r}_{z,b}^2 \rangle^{1/2}$ for $R_{NP} = 20$ and 150 nm is therefore a consequence of the collisions between the crowders and the NP. This conclusion is at first sight counter-intuitive, because isotropic NP/crowders collisions are expected not to alter the mean conformation of the DNA coil. The point, however, is that NP/crowders collisions are precisely NOT isotropic. Indeed, they are less frequent close to the point where the DNA chain is grafted to the NP compared to the other side of the NP. On average, NP/crowders collisions are therefore expected to result in a net force oriented from the center of the NP towards DNA bead $n - 1$, that is, along vector $\mathbf{r}_{n-1} - \mathbf{r}_n$. This conjecture is confirmed by the top plot of Fig. 5, which indicates that the mean value of the projection of force

$$\mathbf{F} = -q^2 \nabla \sum_{j=1}^{C} H(\|\mathbf{r}_n - \mathbf{s}_j\| - R_{NP} - R_C) \tag{18}$$

on vector $\mathbf{r}_{n-1} - \mathbf{r}_n$ is non zero and increases exponentially with $\rho_e$. In addition, especially for the largest NP with radius $R_{NP} = 150$ nm, the DNA coil locates preferentially between origin O, where one end of the DNA chain is grafted, and the NP. As a consequence, vector $\mathbf{r}_{n-1} - \mathbf{r}_n$ points preferentially approximately towards the position of the center of mass G of the DNA coil. This can be checked in the bottom plot of Fig. 5, which displays the evolution with $\rho_e$ of the mean value of the projection of $\mathbf{F}$ on vector $\overrightarrow{OG} - \mathbf{r}_n$. The mean amplitude of the projected force increases again exponentially with $\rho_e$, being of the order of 6 fN at $\rho_e = 0.41$ for $R_{NP} = 1$ nm, 40 fN for $R_{NP} = 20$ nm, and 280 fN for $R_{NP} = 150$ nm. For the sake of comparison, it is reminded that a DNA molecule grafted to a surface is stretched to approximately half its contour length when a constant force of 80 fN is applied to its free end [55]. We furthermore checked that the application at the center of the NPs of a force of 80 fN directed towards the apex of the confinement hemisphere is sufficient to let the NPs reach the apex within about 50 ms (for $R_{NP} = 20$ nm) or 150 ms (for $R_{NP} = 150$ nm) at $\rho_e = 0.41$. Our tentative explanation for the steeper decrease of $\langle R_g^2 \rangle^{1/2}$, $\langle \mathbf{R}_{ee}^2 \rangle^{1/2}$, $\langle \mathbf{r}_\parallel^2 \rangle^{1/2}$ and $\langle \mathbf{r}_{z,b}^2 \rangle^{1/2}$ for $R_{NP} = 20$ and 150 nm compared to the homogeneous DNA chain is therefore that the anisotropy of the collisions between the crowders and NPs with radius $R_{NP} = 20$ or 150 nm results in a force which is sufficiently strong to push back the end of the DNA molecule grafted to the NP towards the center of mass of the DNA coil, thereby contributing to the compaction of the DNA coil.

This is of course a caveat that must be taken into account when using the TPM set-up to investigate the compaction of the DNA coil by globular macromolecular crowders. Indeed, the BD simulations suggest that, if the NP is sufficiently large, then the interactions between the crowders and the NP have a more pronounced effect on the mean lateral displacement of the NP and the compaction of the DNA coil than the interactions between the crowders and the DNA chain. When working in the linear regime, it is consequently probably safer to use the Tethered Fluorophore Motion (TFM) technique, which consists in replacing the NP by a much smaller fluorophore bound to the free end of the DNA chain [4,5], rather than the usual TPM technique.

**Analysis of the regime above $\rho_e = 0.41$**

BD simulations indicate that two different phenomena are likely to take place in a narrow range of crowder concentration above $\rho_e = 0.41$.

On one hand, in the absence of NP ($R_{NP} = 1$ nm), both $\langle R_g^2 \rangle^{1/2}$ and $\langle \mathbf{R}_{ee}^2 \rangle^{1/2}$ decrease by about 45% at $\rho_e = 0.55$ compared to $\rho_e = 0.41$. This sharp compaction is of course reminiscent of the abrupt coil to globule transition induced by above-threshold concentrations of salt and simple neutral polymers [56] or BSA proteins [38,57], which takes place close to the jamming threshold [39]. The DNA concentration measured in compact globules is admittedly much larger than the concentration obtained in BD simulations with $R_{NP} = 1$ nm and $\rho_e = 0.55$, but two features of the model may contribute to limit DNA compaction in the simulations. First, the mean radius of gyration of the DNA coil for $R_{NP} = 1$ nm and $\rho_e = 0.55$, $\langle R_g^2 \rangle^{1/2} \approx 106$ nm, is close to the diameter of the crowders, $2R_C = 80$ nm. This indicates that the compacted DNA coil occupies approximately the place of one crowder in the almost periodic array of crowders that forms close to the jamming threshold (See Fig. 2 of [45] and Fig. 8 of [46]) and suggests that it is not possible to compact the DNA coil further without decreasing the size of the crowders. This hypothesis is supported by the fact that DNA coils with a radius of gyration $\langle R_g^2 \rangle^{1/2} \approx 77$ nm were obtained for the same DNA chain and crowders with $R_C = 7.4$ nm (see Fig. 4 of [46]). Moreover, it has been shown that the condensation of the DNA coil to a globule by anionic nanoparticles is accompanied by the denaturation of the DNA strands [39], which decreases the bending rigidity of the DNA molecule by about one order of magnitude. In order to reproduce this extreme compaction of the DNA coil, it would therefore probably be necessary to take denaturation properly into account in the coarse–grained model.

On the other hand, BD simulations performed with $R_{NP} = 150$ nm and $\rho_e = 0.48$ suggest that NPs with radius $R_{NP} = 150$ nm remain in the vicinity of the bottom plane as soon as they are first brought in its neighborhood by random thermal motion, in agreement with theoretical predictions [52]. Most probably, if crowders of the size of BSA proteins ($\approx 4.5$ nm [51]) were used, then NPs with $R_{NP} = 20$ nm would also remain most of the time close to the bottom plane.

The fact that these two mechanisms take place in the same narrow interval of crowder concentration points towards a further potential experimental difficulty. Indeed, streptavidin-coated NPs tend to be attracted by, and adhere to, the anti-digoxigenin-coated spots which cover the bottom coverslip of most TPM experimental set-ups. At crowder concentrations where NPs are maintained most of the time in the vicinity of the bottom surface, most of them are probably trapped and immobilized by the coating. The problem is that it is actually difficult to discriminate experimentally between NPs which are immobilized because they adhere to the coating and NPs which are maintained at a short distance of the bottom surface by collapsed DNA molecules. In order to lift this uncertainty, one may eventually have recourse to more efficient sample preparation techniques, which do not involve hydrophilic patches to which streptavidin-coated NPs may adhere [58].

**Range of DNA concentrations available with the TPM set-up**

In connection with the previous point, BD simulations suggest that the sole addition of globular crowders in a standard TPM set-up will probably not be sufficient to investigate the range of DNA concentration that prevails in prokaryotic cells. Indeed, in the linear regime ($\rho_e \leq 0.41$) the compaction of the DNA coil by globular crowders does not exceed a few percents and remains quite modest compared to the one that prevails in living prokaryotic cells. Indeed, the global density of DNA base pairs in living bacteria ($\approx 5$ mM) corresponds to that of a 7500 bp DNA molecule enclosed in a sphere of radius $\approx 84$ nm and having a radius of gyration of $\approx 65$ nm. Moreover, macromolecular crowders, DNA-bridging proteins and supercoiling contribute to compacting the genomic DNA into an even denser object called the nucleoid [33], which occupies only a fraction of the cell. This means that the 65 nm value, which is shown as a horizontal gray dot-dashed line in the top plot of Fig. 2, should be considered as an upper limit for the 7500 bp DNA to be as compact as in living cells. Actually, the radius of gyration of the DNA coil at $\rho_e = 0.41$ is about 3 times larger than this limit and corresponds to a concentration of DNA about 30 times smaller.

In contrast, the DNA concentration in compact globules formed by above-threshold concentrations of monovalent or divalent salt and simple neutral polymers [56], or small trivalent or tetravalent polycations like spermidine and spermine [59-61], is much larger than the DNA concentration in living cells. For example, the concentration of base pairs is of the order of 200-300 mg/ml, that is, about 0.3-0.45 M, in globules formed with increasing concentrations of PEG and salt [62]. This is about 2 orders of magnitude larger than in living cells and would correspond to a radius of gyration of about 17 nm for a 7500 bp DNA having the shape of a spherical globule. The point is that the coil to globule transition corresponds to a phase transition, so that all DNA concentrations comprised between that of a slightly compacted coil and that of a dense globule cannot be attained with these methods. In order to achieve DNA concentrations close to in vivo ones, one must instead use methods which lead to a gradual compaction of the DNA coil. For example, one could think of adding to the injected solvent either long (≈100 monomers) poly-L-lysine molecules [63,64] or DNA-bridging proteins, like H-NS [65]. Such proteins are indeed able to compact substantially the DNA coil and compaction ratios due to globular crowders and DNA-bridging proteins are additive over large concentration ranges [46]. Still, upon addition of poly-L-lysine or DNA-bridging proteins, one would still have to face a problem related to the rigidity of the DNA molecule. Indeed, the radius of gyration of the coil formed by a 7500 bp DNA molecule at concentrations close to in vivo ones (less than 65 nm) is of the same order of magnitude as the persistence length of the DNA molecule ($\xi = 50$ nm), so that the bending rigidity of the DNA molecule would probably oppose significantly the compaction of the DNA coil by globular crowders and DNA-bridging molecules. This problem could in turn be somewhat alleviated by working with longer DNA molecules, of the order of 20000 bp instead of 7500 bp, which however lie at the limit of today's TPM feasibility.

Considering all the points raised in the present study, a promising method for applying the TPM technique to dense media with DNA concentration close to in vivo may perhaps consist in focusing on the aggregation of multiple DNA molecules instead of the condensation of a single one and preparing ad-hoc substrates. For example, one may use soft nano-lithography to graft ≈7500 bp DNA molecules on non-hydrophilic patches of size 100 nm × 100 nm, which would result in a DNA layer of thickness ≈400 nm and base pair concentration ≈5 mM. One would furthermore attach a fluorophore at the free end of the DNA molecules located on a lattice of period ≈1 μm. By recording the motion of these fluorophores with a htTPM set-up, one should be able to characterize the compaction of the DNA layer upon addition of increasing concentrations of globular crowders and/or DNA bridging proteins. While technically feasible with today available technologies, this project would probably still require some developments in order to firmly relate the amplitude of motion of the fluorophores to the properties of the DNA layer.

**CONCLUSION**

In this work, extensive BD simulations were launched to estimate whether TPM is a suitable technique for investigating crowded media, with a focus on the compaction of the DNA molecule by globular crowders. The simulations highlight several effects not observed in dilute media, which render the interpretation of experimental signals potentially more complex. For example, the interactions between the crowders and the NP may affect the mean lateral displacement of the NP and the compaction of the DNA coil more strongly than the interactions between the crowders and the DNA molecule. Moreover, it may be difficult to distinguish between NPs which are brought in the neighborhood of the bottom glass by entropic effects and adhere to the coating and NPs which are maintained at a short distance of the bottom surface by collapsed DNA molecules. These issues can hopefully be solved technically by using small fluorophores instead of the larger NPs and having recourse to sample preparation techniques that do not involve hydrophilic patches. Simulations furthermore suggest that DNA concentrations close to in vivo ones could be difficult to achieve with a standard TPM set-up. The solution may consist in attaching the fluorophores to a small portion of the DNA molecules forming a DNA brush.

It is hoped that this work will trigger more work, both experimental and theoretical.

**Declaration of Interests**

The author reports there are no competing interests to declare.

**Acknowledgements**

This work was supported by the MITI CNRS under Grant Modélisation du Vivant 2021 and 2022 "StatPhysProk".

**REFERENCES**

1. D. A. Schafer, J. Gelles, M. P. Sheetz, R. Landick. Transcription by single molecules of RNA polymerase observed by light microscopy. *Nature* 1991, 352, 444–448.

2. H. Yin, R. Landick, J. Gelles. Tethered particle motion method for studying transcript elongation by a single RNA polymerase molecule. *Biophys. J.* 1994, 67, 2468–2478.

3. T. Plénat, C. Tardin, P. Rousseau, L. Salomé. High-throughput single-molecule analysis of DNA-protein interactions by tethered particle motion. *Nucleic Acids Res.* 2012, 40, e89.

4. J. N. M. Pinkney, P. Zawadzki, J. Mazuryk, L. K. Arciszewska, D. J. Sherratt, A. N. Kapanidis. Capturing reaction paths and intermediates in Cre-loxP recombination using single-molecule fluorescence, *Proc. Natl. Acad. Sci. USA* 2012, 109, 20871–20876.

5. P. F. J. May, J. N. M. Pinkney, P. Zawadzki, G. W. Evans, D. J. Sherratt, A. N. Kapanidis. Tethered fluorophore motion: studying large DNA conformational changes by single-fluorophore imaging. *Biophys. J.* 2014, 107, 1205–1216.

6. H. Yin, I. Artsimovitch, R. Landick, J. Gelles. Nonequilibrium mechanism of transcription termination from observations of single RNA polymerase molecules. *Proc. Natl. Acad. Sci. USA* 1999, 96, 13124–13129.

7. C. Chung, H. W. Li. Direct observation of RecBCD helicase as single-stranded DNA translocases. *J. Am. Chem. Soc.* 2013, 135, 8920–8925.

8. C. Dennis, A. Fedorov, E. Käs, L. Salomé, M. Grigoriev. RuvAB-directed branch migration of individual Holliday junctions is impeded by sequence heterology. *EMBO J.* 2004, 23, 2413–2422.

9. F. Vanzi, C. Broggio, L. Sacconi, F. S. Pavone. Lac repressor hinge flexibility and DNA looping: single molecule kinetics by tethered particle motion. *Nucleic Acids Res.* 2006, 34, 3409–3420.

10. J. Q. Boedicker, H. G. Garcia, S. Johnson, R. Phillips. DNA sequence-dependent mechanics and protein-assisted bending in repressor-mediated loop formation. *Phys. Biol.* 2013, 10, 066005.

11.     S. Johnson, J. W. van de Meent, R. Phillips, C. H. Wiggins, M. Lindén. Multiple LacI mediated loops revealed by Bayesian statistics and tethered particle motion. *Nucleic Acids Res.* 2014, 42, 10265–10277.

12.     Y. Y. Biton, S. Kumar, D. Dunlap, D. Swigon. Lac repressor mediated DNA looping: Monte Carlo simulation of constrained DNA molecules complemented with current experimental results. *PloS One* 2014, 9, e92475.

13.     N. Laurens, S. R. Bellamy, A. F. Harms, Y. S. Kovacheva, S. E. Halford, G. J. Wuite. Dissecting protein-induced DNA looping dynamics in real time. *Nucleic Acids Res.* 2009, 37, 5454–5464.

14.     N. Pouget, C. Turlan, N. Destainville, L. Salomé, M. Chandler. IS911 transpososome assembly as analysed by tethered particle motion. *Nucleic Acids Res.* 2006, 34, 4313-4323.

15.     H. F. Fan, C. H. Ma, M. Jayaram. Single-molecule tethered particle motion: stepwise analyses of site-specific DNA recombination. *Micromachines* 2018, 9, 216.

16.     F. Fournes, E. Crozat, C. Pages, C. Tardin, L. Salomé, F. Cornet, P. Rousseau. FtsK translocation permits discrimination between an endogenous and an imported Xer/dif recombination complex. *Proc. Natl. Acad. Sci. USA* 2016, 113, 7882–7887.

17.     P. Zawadzki, P. F. J. May, R. A. Baker, J. N. M. Pinkney, A. N. Kapanidis, D. J. Sherratt, L. K. Arciszewska. Conformational transitions during FtsK translocase activation of individual XerCDedif recombination complexes. *Proc. Natl. Acad. Sci. USA* 2013, 110, 17302-17307.

18.     G. Nir, M. Lindner, H. R. C. Dietrich, O. Girshevitz, C. E. Vorgias, Y. Garini. HU protein induces incoherent DNA persistence length. *Biophys. J.* 2011, 100, 784-790.

19.     S. F. Tolic-Norrelykke, M. B. Rasmussen, F. S. Pavone, K. Berg-Sorensen, L. B. Oddershede. Stepwise bending of DNA by a single TATA-box binding protein. *Biophys. J.* 2006, 90, 3694–3703.

20.     H. F. Fan, Z. N. Liu, S. Y. Chow, Y. H. Lu, H. Li. Histone chaperone-mediated nucleosome assembly process, *PLoS One* 2015, 10, e0115007.

21.     H. Wang, I. B. Dodd, D. D. Dunlap, K. E. Shearwin, L. Finzi. Single molecule analysis of DNA wrapping and looping by a circular 14mer wheel of the bacteriophage 186 CI repressor. *Nucleic Acids Res.* 2013, 41, 5746–5756.

22. A. H. Mack, D. J. Schlingman, R. D. Salinas, L. Regan, S. G. Mochrie. Condensation transition and forced unravelling of DNA-histone H1 torroids: A multi-state free energy landscape. *J. Phys. Condens. Matter* 2015, 27, 064106.

23. R. F. Guerra, L. Imperadori, R. Mantovani, D. D. Dunlap, L. Finzi. DNA compaction by the nuclear factor-Y. *Biophys. J.* 2007, 93, 176–182.

24. K. Norregaard, M. Andersson, K. Sneppen, P. E. Nielsen, S. Brown, L. B. Oddershede. DNA supercoiling enhances cooperativity and efficiency of an epigenetic switch. *Proc. Natl. Acad. Sci. USA* 2013, 110, 17386-17391.

25. S.-N. Lin, G. J. L. Wuite, R. T. Dame. Effect of different crowding agents on the architectural properties of the bacterial nucleoid-associated protein HU. *Int. J. Mol. Sci.* 2020, 21, 9553.

26. M. Manghi, C. Tardin, J. Baglio, P. Rousseau, L. Salomé, N. Destainville. Probing DNA conformational changes with high temporal resolution by tethered particle motion. *Phys. Biol.* 2010, 7, 046003.

27. S. Cayley, B. A. Lewis, H. J. Guttman, M. T. Record. Characterization of the cytoplasm of Escherichia coli K-12 as a function of external osmolarity. Implications for protein-DNA interactions in vivo. *J. Mol. Biol.* 1991, 222, 281–300.

28. S. B. Zimmerman, S. O. Trach. Estimation of macromolecule concentrations and excluded volume effects for the cytoplasm of Escherichia coli. *J. Mol. Biol.* 1991, 222, 599–620.

29. P. Chaudhuri, L. Berthier, S. Sastry. Jamming transitions in amorphous packings of frictionless spheres occur over a continuous range of volume fractions. *Phys. Rev. Lett.* 2010, 104, 165701.

30. N. Bellotto, J. Agudo-Canalejo, R. Colin, R. Golestanian, G. Malengo, V. Sourjik. Dependence of diffusion in Escherichia coli cytoplasm on protein size, environmental conditions, and cell growth. *eLife* 2022, 11, e82654.

31. S. B. Zimmerman. Shape and compaction of Escherichia coli nucleoids. *J. Struct. Biol.* 2006, 156, 255-261.

32. R. de Vries. DNA condensation in bacteria: Interplay between macromolecular crowding and nucleoid proteins. *Biochimie* 2010, 92, 1715–1721.

33. V. G. Benza, B. Bassetti, K. D. Dorfman, V. F. Scolari, K. Bromek, P. Cicuta, M. C. Lagomarsino. Physical descriptions of the bacterial nucleoid at large scales, and their biological implications. *Rep. Prog. Phys.* 2012, 75, 076602.

34. M. Joyeux. Compaction of bacterial genomic DNA: Clarifying the concepts. *J. Phys. Condens. Matter* 2015, 27, 383001.

35. M. Joyeux. In vivo compaction dynamics of bacterial DNA: A fingerprint of DNA/RNA demixing ? *Curr. Opin. Colloid. Interface Sci.* 2016, 26, 17-27.

36. M. Castelnovo, W. M. Gelbart. Semiflexible chain condensation by neutral depleting agents: Role of correlations between depletants. *Macromolecules* 2004, 37, 3510-3517.

37. R. de Vries. Depletion-induced instability in protein-DNA mixtures: Influence of protein charge and size. *J. Chem. Phys.* 2006, 125, 014905.

38. M. K. Krotova, V. V. Vasilevskaya, N. Makita, K. Yoshikawa, A. R. Khokhlov. DNA compaction in a crowded environment with negatively charged proteins. *Phys. Rev. Lett.* 2010, 105, 128302.

39. A. Zinchenko, K. Tsumoto, S. Murata, K. Yoshikawa. Crowding by anionic nanoparticles causes DNA double-strand instability and compaction. *J. Phys. Chem. B* 2014, 118, 1256-1262.

40. S. Bakshi, H. Choi, J. Mondal, J. C. Weisshaar. Time-dependent effects of transcription- and translation-halting drugs on the spatial distributions of the Escherichia coli chromosome and ribosomes. *Mol. Microbiol.* 2014, 94, 871-887.

41. S. Bakshi, H. Choi, J. C. Weisshaar. The spatial biology of transcription and translation in rapidly growing Escherichia coli. *Front. Microbiol.* 2015, 6, 636.

42. M. Joyeux. Coarse-grained model of the demixing of DNA and non-binding globular macromolecules. *J. Phys. Chem. B* 2017, 121, 6351-6358.

43. M. Joyeux. A segregative phase separation scenario of the formation of the bacterial nucleoid. *Soft Matter* 2018, 14, 7368-7381.

44. M. Joyeux. Bacterial nucleoid: Interplay of DNA demixing and supercoiling. *Biophys. J.* 2019, 118, 2141-2150.

45. M. Joyeux. Preferential localization of the bacterial nucleoid. *Microorganisms* 2019, 7, 204.

46. M. Joyeux. Organization of the bacterial nucleoid by DNA-bridging proteins and globular crowders. *Front. Microbiol.* 2023, 14, 1116776.

47. H. Jian, A. Vologodskii, T. Schlick. A combined wormlike-chain and bead model for dynamic simulations of long linear DNA. *J. Comp. Phys.* 1997, 136, 168-179.

48. G. S. Manning. Limiting laws and counterion condensation in polyelectrolyte solutions. I. Colligative properties. *J. Chem. Phys.* 1969, 51, 924-933.

49. F. Oosawa. Polyelectrolytes. Marcel Dekker, New York, 1971.

50. K. A. Rubinson, S. Krueger. Poly(ethylene glycol)s 2000-8000 in water may be planar: A small-angle neutron scattering (SANS) structure study. *Polymer* 2009, 50, 4852-4858.

51. U. Bohme, U. Scheler. Effective charge of bovine serum albumin determined by electrophoresis NMR. *Chem. Phys. Lett.* 2007, 435, 342–345.

52. R. Roth, B. Götzelmann, S. Dietrich. Depletion forces near curved surfaces. *Phys. Rev. Lett.* 1999, 83, 448-451.

53. I. Teraoka. Polymer Solutions: An Introduction to Physical Properties. Wiley, New-York, 2002.

54. D. E. Segall, P. C. Nelson, R. Phillips. Volume-exclusion effects in tethered-particle experiments: Bead size matters. *Phys. Rev. Lett.* 2006, 96, 088306.

55. J. F. Marko, E. D. Siggia. Stretching DNA. *Macromolecules* 1995, 28, 8759-8770.

56. L. S. Lerman. A transition to a compact form of DNA in polymer solutions. *Proc. Natl. Acad. Sci. USA* 1971, 68, 1886-1890.

57. K. Yoshikawa, S. Hirota, N. Makita, Y. Yoshikawa. Compaction of DNA induced by like-charge protein: Opposite salt-effect against the polymer-salt-induced condensation with neutral polymer. *J. Phys. Chem. Lett.* 2010, 1, 1763-1766.

58. L. Jin, L. Kou, Y. Zeng, C. Hu, X. Hu. Sample preparation method to improve the efficiency of high-throughput single-molecule force spectroscopy. *Biophys. Rep.* 2019, 5, 176–183.

59. R. W. Wilson, V. A. Bloomfield. Counterion-induced condensation of deoxyribonucleic acid. A light scattering study. *Biochemistry* 1979, 18, 2192-2196.

60. J. Pelta, F. Livolant, J.-L. Sikorav. DNA aggregation induced by polyamines and cobalthexamine. *J. Biol. Chem.* 1996, 271, 5656–5662.

61. Y. Burak, G. Ariel, D. Andelman. Onset of DNA aggregation in presence of monovalent and multivalent counterions. Biophys. J. 2003, 85, 2100-2110.

62 T. Maniatis, J. H. Venable, L. S. Lerman. The structure of ψ DNA. *J. Mol. Biol.* 1974, 84, 37-64.

63. J. DeRouchey, R. R. Netz, J. O. Rädler. Structural investigations of DNA-polycation complexes. *Eur. Phys. J. E* 2005, 16, 17–28.

64. T. Akitaya, A. Seno, T. Nakai, N. Hazemoto, S. Murata, K. Yoshikawa. Weak interaction induces an ON/OFF switch, whereas strong interaction causes gradual change: Folding transition of a long duplex DNA chain by poly-L-lysine. *Biomacromolecules* 2007, 8, 273-278.

65. L. Qin, A. M. Erkelens, F. Ben Bdira, R. T. Dame. The architects of bacterial DNA bridges: A structurally and functionally conserved family of proteins. *Open Biology* 2019, 9, 190223.

**Table 1.** Coefficients $A$ and $B$ of Eq. (13) and relative decrease $\Delta$ of each quantity between $\rho_e = 0$ and $\rho_e = 0.41$ computed according to Eq. (14).

|  |  | $A$ (nm) | $B$ (nm) | $\Delta$ (%) |
|---|---|---|---|---|
| $\langle R_g^2 \rangle^{1/2}$ | $R_{NP} = 1$ nm | 204.1 ± 3.9 | -26.1 ± 15.4 | -5.2 |
|  | $R_{NP} = 20$ nm | 207.9 ± 1.6 | -87.0 ± 6.3 | -17.2 |
|  | $R_{NP} = 150$ nm | 218.1 ± 4.2 | -93.0 ± 16.5 | -17.5 |
| $\langle \mathbf{R}_{ee}^2 \rangle^{1/2}$ | $R_{NP} = 1$ nm | 544.3 ± 11.3 | -16.5 ± 44.3 | -1.2 |
|  | $R_{NP} = 20$ nm | 562.5 ± 3.5 | -253.3 ± 13.6 | -18.5 |
|  | $R_{NP} = 150$ nm | 635.7 ± 19.2 | -326.9 ± 75.2 | -21.1 |
| $\langle \mathbf{r}_\parallel^2 \rangle^{1/2}$ | $R_{NP} = 1$ nm | 410.3 ± 2.1 | -17.0 ± 8.1 | -1.7 |
|  | $R_{NP} = 20$ nm | 409.6 ± 24.8 | -143.0 ± 97.3 | -14.3 |
|  | $R_{NP} = 150$ nm | 501.7 ± 32.9 | -306.3 ± 128.9 | -25.0 |
| $\langle \mathbf{r}_{z,b}^2 \rangle^{1/2}$ | $R_{NP} = 1$ nm | 356.5 ± 14.9 | -5.2 ± 58.5 | -0.6 |
|  | $R_{NP} = 20$ nm | 368.8 ± 27.2 | -216.8 ± 106.9 | -24.1 |
|  | $R_{NP} = 150$ nm | 351.3 ± 9.6 | -193.9 ± 37.8 | -22.6 |

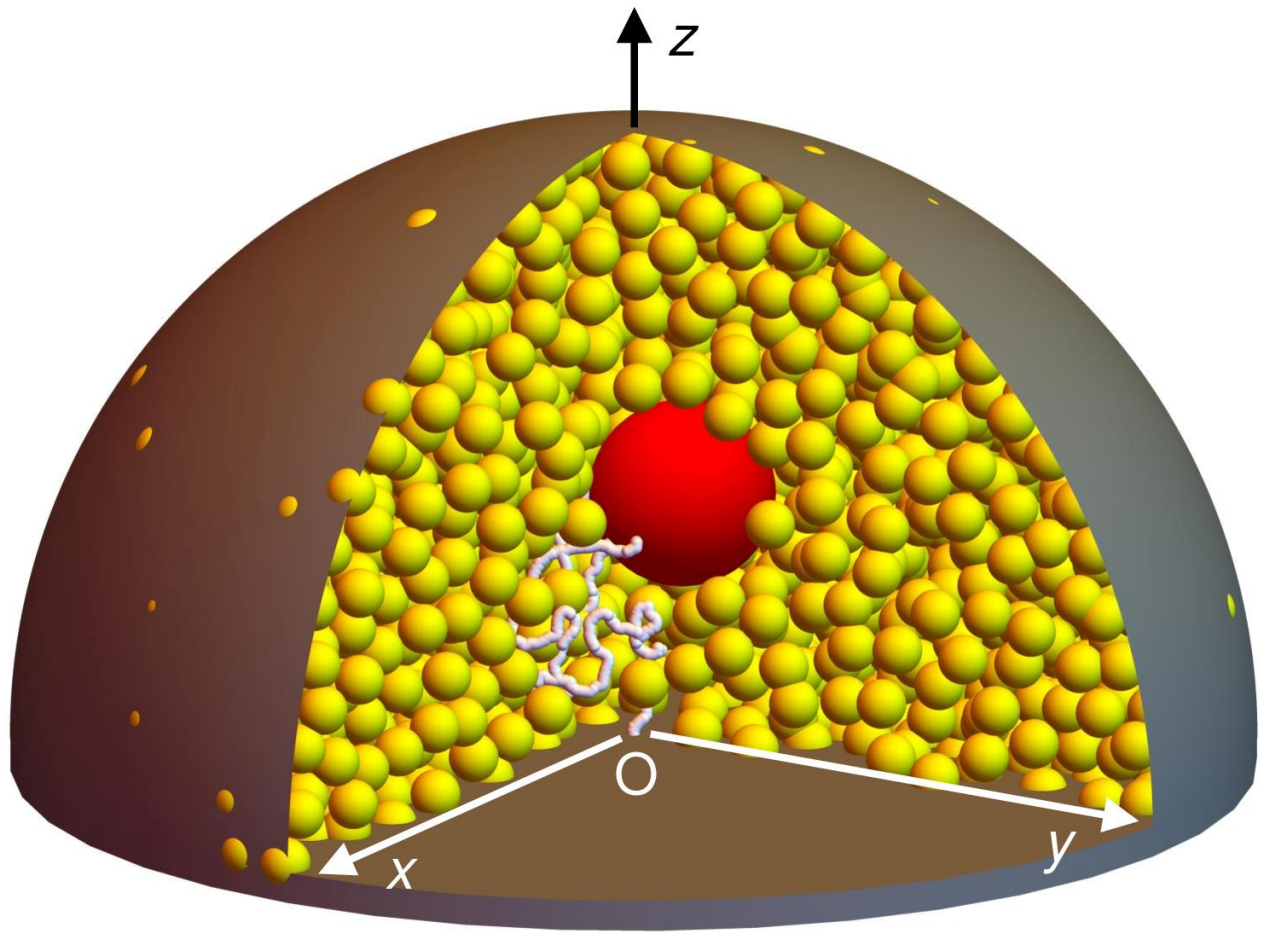

**Figure 1**. Snapshot extracted from a BD simulation with $C = 3000$ crowders and $R_{\text{NP}} = 150$ nm. The white chain represents the DNA molecule, the red ball the NP and yellow balls the crowders. The radius of DNA beads has been multiplied by 10 for the sake of clarity. One fourth of the confinement hemisphere and of the crowders have been removed, in order for the DNA chain and the NP to be seen more clearly.

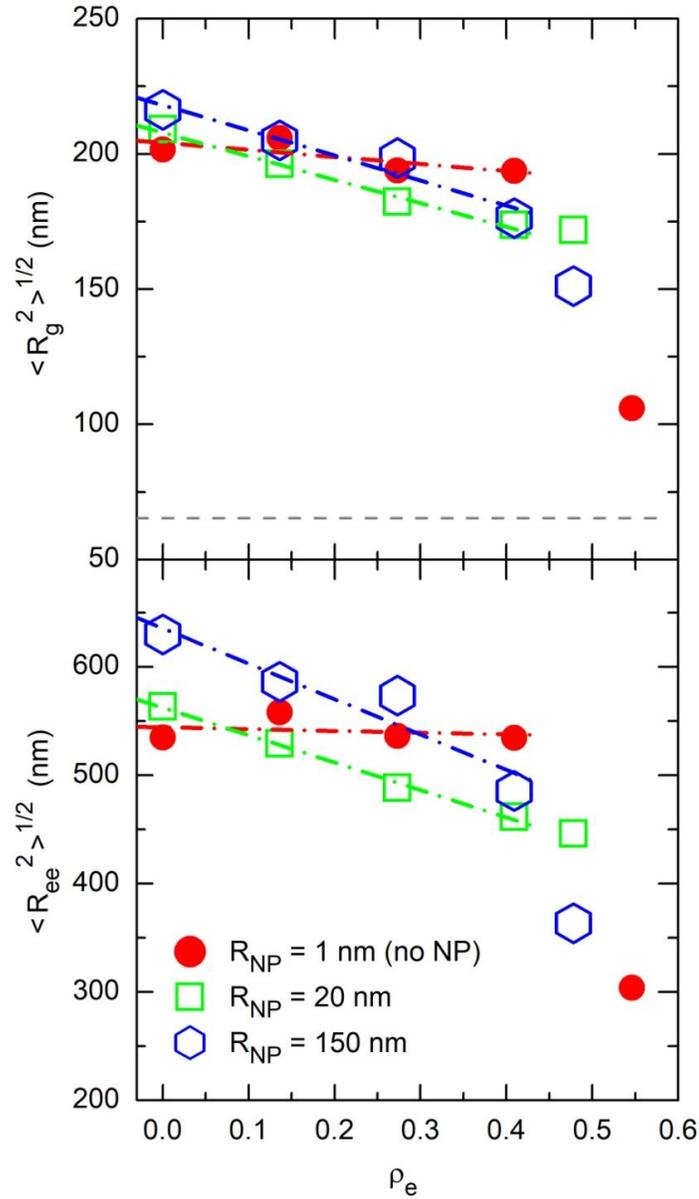

**Figure 2**. Evolution of $\langle R_g^2 \rangle^{1/2}$ (top plot) and $\langle \mathbf{R}_{ee}^2 \rangle^{1/2}$ (bottom plot) as a function of $\rho_e$. Red dots, green squares and blue hexagons represent the results of BD simulations performed with $R_{NP} = 1, 20$ and 150 nm, respectively ($R_{NP} = 1$ nm corresponds to a homogenous DNA chain with no terminal NP). Dot-dashed lines represent the result of linear fits against the four points with $\rho_e \leq 0.41$ for each value of $R_{NP}$. The dashed horizontal line at $R_g = 65$ nm indicates the radius of gyration that the 7500 bp DNA molecule would have if it were as dense as in a decompacted bacterial nucleoid.

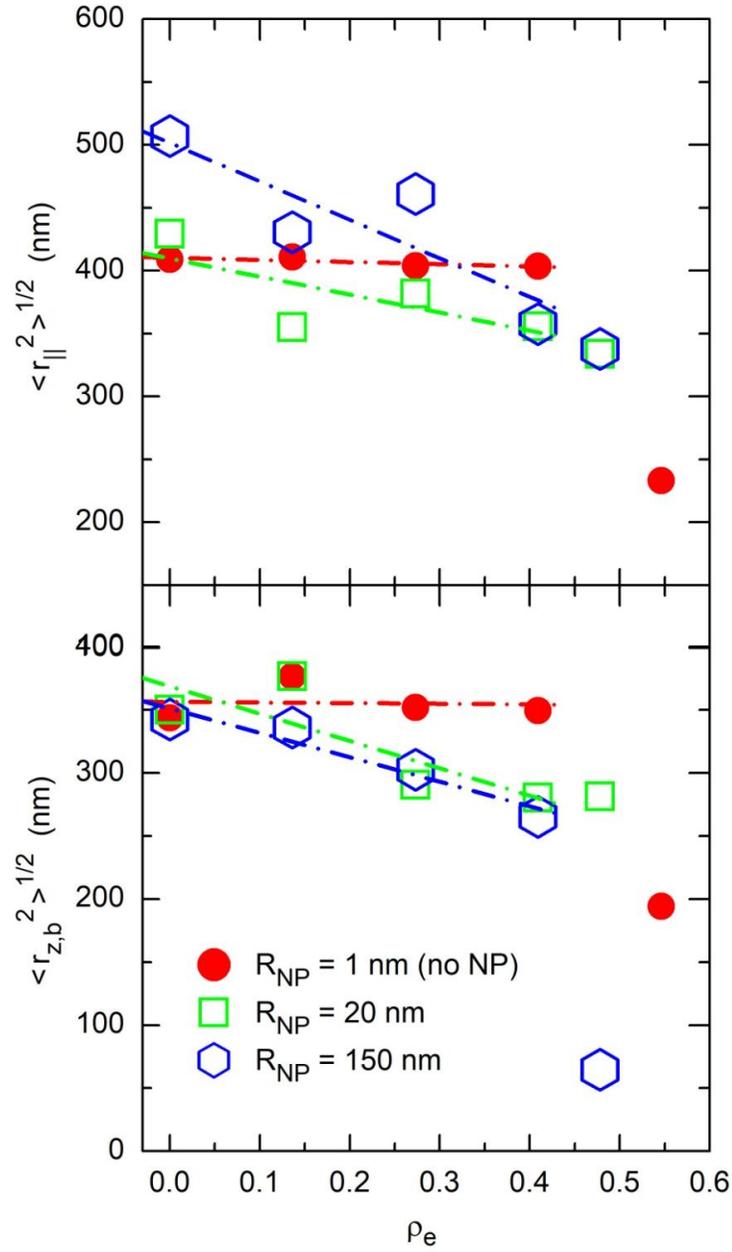

**Figure 3**. Evolution of $\langle r_{\parallel}^2 \rangle^{1/2}$ (top plot) and $\langle r_{z,b}^2 \rangle^{1/2}$ (bottom plot) as a function of $\rho_e$. Red dots, green squares and blue hexagons represent the results of BD simulations performed with $R_{NP} = 1$, 20 and 150 nm, respectively ($R_{NP} = 1$ nm corresponds to a homogenous DNA chain with no terminal NP). Dot-dashed lines represent the result of linear fits against the four points with $\rho_e \leq 0.41$ for each value of $R_{NP}$.

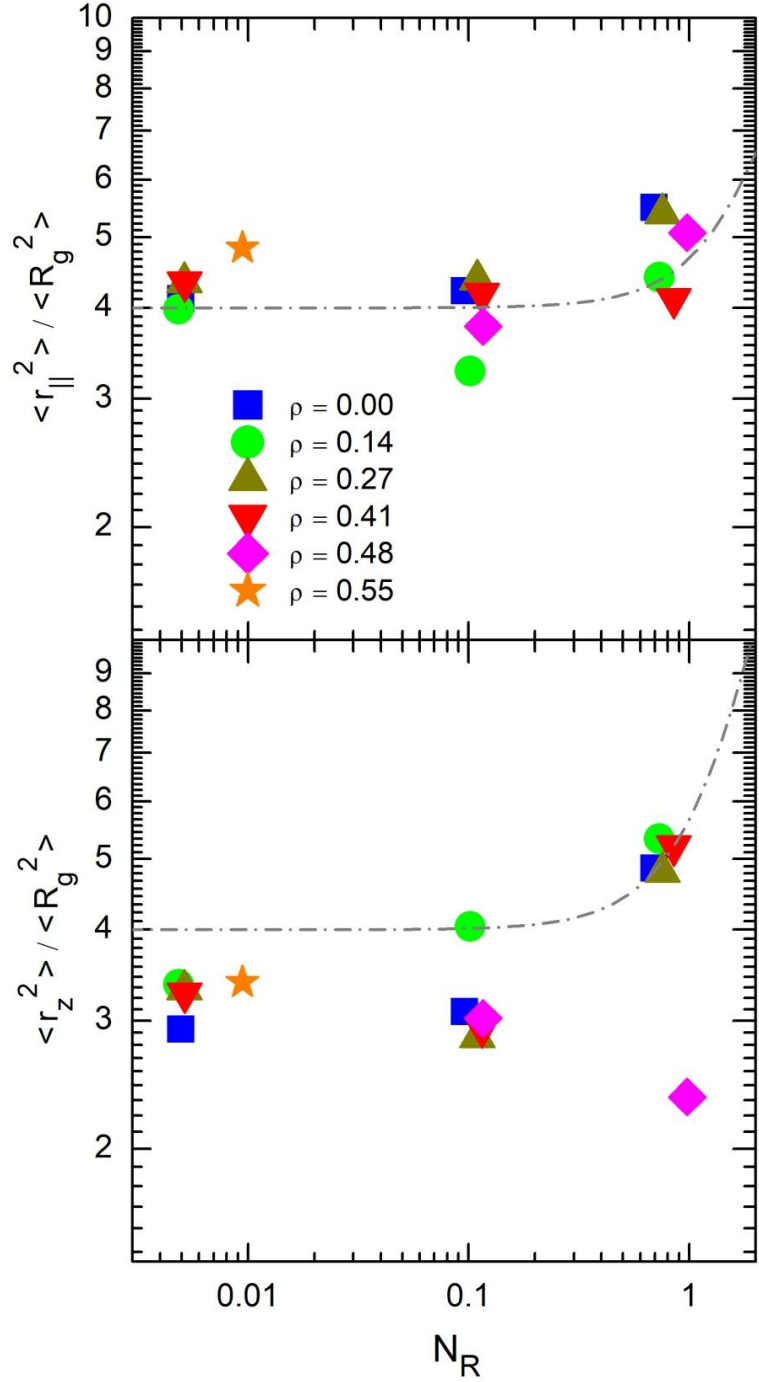

**Figure 4**. Log-log plots of the evolution of $\langle \mathbf{r}_{\parallel}^2 \rangle / \langle R_g^2 \rangle$ (top plot) and $\langle \mathbf{r}_z^2 \rangle / \langle R_g^2 \rangle$ (bottom plot) as a function of $N_R$. Symbols represent the results obtained from BD simulations and gray dot-dashed lines the theoretical estimates in the right-hand sides of Eqs. (16) and (17).

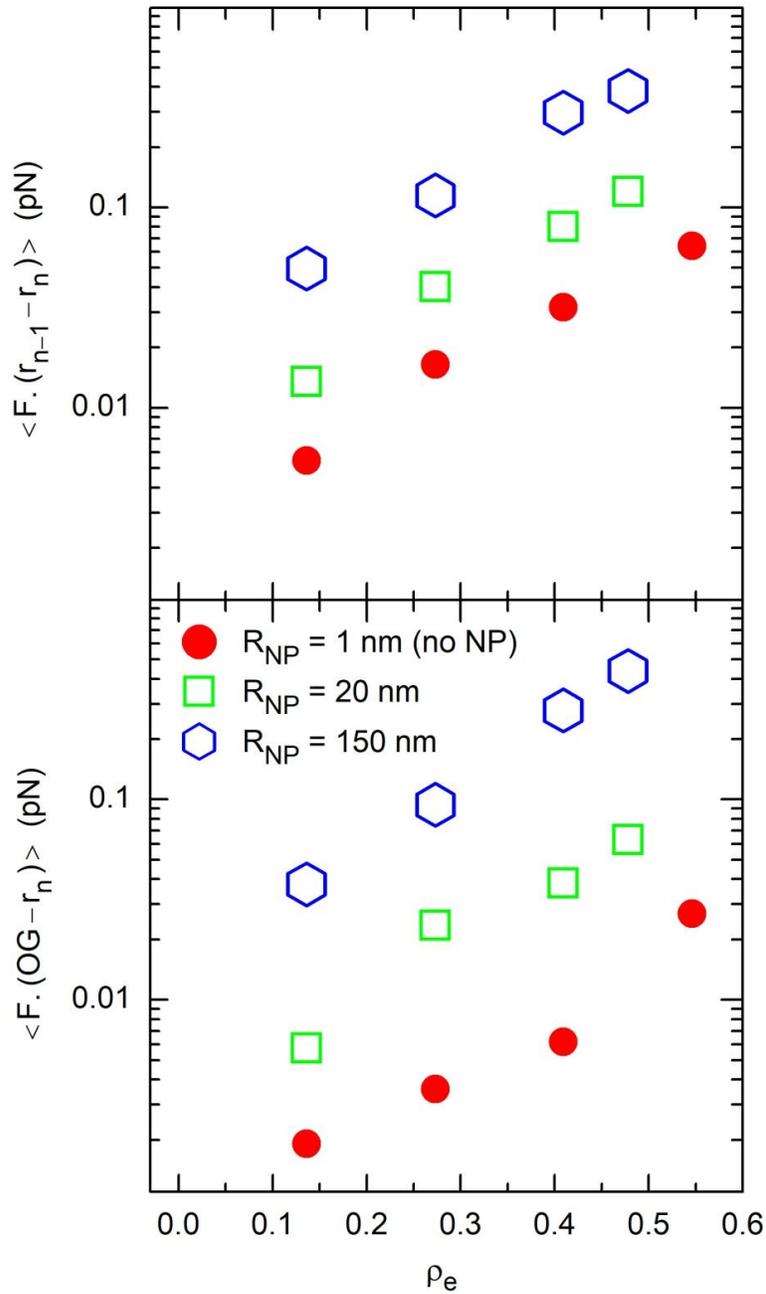

**Figure 5**. Evolution of $\langle \mathbf{F} \cdot (\mathbf{r}_{n-1} - \mathbf{r}_n) \rangle$ (top plot) and $\langle \mathbf{F} \cdot (\overrightarrow{OG} - \mathbf{r}_n) \rangle$ (bottom plot) as a function of $\rho_e$. Red dots, green squares and blue hexagons represent the results of BD simulations performed with $R_{NP} = 1$, 20 and 150 nm, respectively ($R_{NP} = 1$ nm corresponds to a homogenous DNA chain with no terminal NP).

**TOC Graphic**

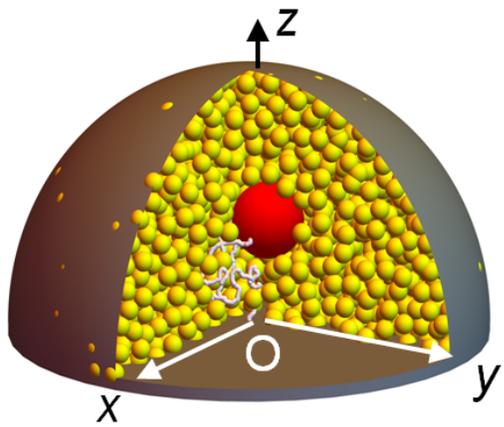
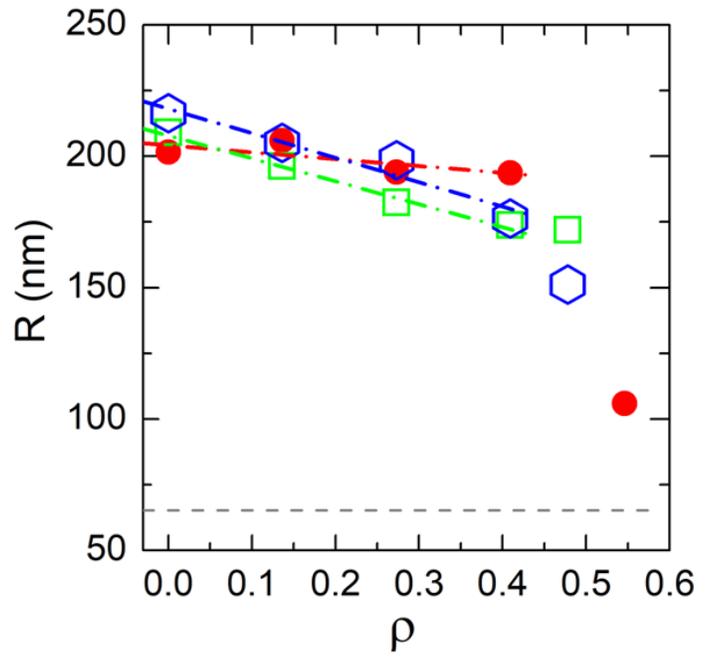